\documentclass[12pt]{article}

\usepackage{amsfonts,amssymb,amsbsy,latexsym,amsmath}

\newcommand{\be}{\begin{equation}}
\newcommand{\ee}{\end{equation}}
\newcommand{\ba}{\begin{eqnarray}}
\newcommand{\ea}{\end{eqnarray}}
\newcommand{\pa}{\partial}
\let\f\frac
\newcommand{\s}{\sqrt}

\newcommand{\ds}{\displaystyle}
\newcommand{\ed}{\end{document}}

\begin{document}

\begin{center}
{\large\bf HOW WERE  THE HILBERT--EINSTEIN}\\

\vspace*{0.2cm}
{\large\bf EQUATIONS DISCOVERED?}
\\

\vspace*{0.4cm}
{\bf A.A. Logunov, M.A.Mestvirishvili, V.A. Petrov\footnote{e-mail:
Vladimir.Petrov@ihep.ru}}

\vspace*{0.4cm}
{\it Division of Theoretical Physics,\\ Institute for High Energy
Physics,\\
142281, Protvino, Moscow Region, Russian Federation }

\end{center}

\begin{abstract}
The pathways along which A.~Einstein and D.~Hilbert independently
came to the gravitational field equations are traced. Some of the
papers that assert a point of view on the history of the derivation
of the gravitational field equations {\it ``that radically differs
from the standard point of view''} are critically analyzed. It is
shown that the conclusions drawn in these papers are completely
groundless.  
\end{abstract}

\section*{Introduction}

Since the studies by J.~Earman and C.~Glymour [1] it became clear
that the equations of A.~Einstein's general relativity were
discovered
almost simultaneously, but with different methods, by D.~Hilbert and
A.~Einstein.

In 1997 an article appeared in the journal ``Science'' under the
title {\sf``Belated Decision in the Hilbert--Einstein Priority
Dispute''}~[2], the  authors of which
claim that {\it ``...knowledge of Einstein's result may have been
crucial to Hilbert's introduction of the trace term into his field
equations''}. On this ground they push forward their
point of view {\it ``that radically differs from the standard point
of
view''} and which is  exposed in a many-page ref.[3].

According to the standard point of view Einstein and Hilbert,
independently  of each other and in different ways, discovered the
gravitational field equations. The same question was the subject of
the paper [4]. 

What is the question? In the Einstein paper [5] the gravitational
field equations are given:
\[
\sqrt{-g}R_{\mu\nu}=-{\varkappa}\left(T_{\mu\nu}-\f{1}{2}g_{\mu\nu}
T\right)\,
,
\]
where, as usual, $ g_{\mu\nu}$  is a metric tensor;  
$R_{\mu\nu}$  is the  Ricci tensor,
$\varkappa$ stands for the gravitational coupling constant, 
$T_{\mu\nu}$  is the energy-momentum tensor density for matter,  
$T$ is the trace of $T_{\mu\nu}$: 
\[ 
T=g^{\mu\nu}T_{\mu\nu}\, . 
\]
The authors of the paper [2] assert that  Hilbert,  when having
taken 
knowledge of these equations and  having seen the ``trace term''
$\bigl(\ds\f{1}{2}g_{\mu\nu}T\bigr)$, would be also ``introduced'' 
into his equations [6], 
\be
\sqrt{g}\left(R_{\mu\nu}-\f{1}{2}g_{\mu\nu}R\right)
=-\f{\pa \sqrt{g}\,L}{\pa g^{\mu\nu}}, 
\ee
the trace term  (in this case $\ds\f{1}{2}g_{\mu\nu}R$\, ,
where the trace 
$R=g^{\mu\nu}R_{\mu\nu}$).  

Let us see in what field equation Hilbert needed, according to the
authors [2], to ``introduce the trace term''. The authors of ref.
[2] do not take into account  that in  the Hilbert approach nothing
can be
``introduced'' because everything is exactly defined by  the world
function (Lagrangean). 
\[
H=R+L \, ,
\]
discovered by Hilbert, which plays a key role for derivation of the
gravitational equations in the framework of the least action
principle. 

The authors of [2] produced their discovery  when they took
knowledge of
the proofs of the Hilbert paper (in which, by the way,  some
parts are missed. See [7], where, in particular, the remained parts
of the proofs are reproduced) and saw that the gravitational field
equations were presented there in the form  of the variational
derivative of 
$[\sqrt{g}R]$ in $g^{\mu\nu}$ 
\be
\f{\pa \sqrt{g}R}{\pa g^{\mu\nu}}
-\pa_k \f{\pa \sqrt{g}R}{\pa g_k^{\mu\nu}}
+\pa_k \pa_\ell \f{\pa \sqrt{g}R}{\pa g_{k\ell}^{\mu\nu}}
=-\f{\pa \sqrt{g}L}{\pa g^{\mu\nu}},
\ee 
but not in the form (1). Thereof they draw their conclusion that
Hilbert did not derived the gravitation equations in the form (1). 

{\bf But if  even everything were so}, then at any rate Hilbert
needed
nothing to ``introduce'' in addition because Eq.(2) turns exactly
into Eq.(1) after some quite trivial  calculations. 
{ \bf Things, however, go not in such a way as authors of [2] wrote.
}
In order to show that the statement by the authors of [2] has no
serious grounds we have to give an account of the basics of 
D.~Hilbert's work (see Section~1).

On  the basis of the idea of equivalence of acceleration and gravity
Einstein in the joint article [8] with M.~Grossmann  in 1913
identified the gravitational field and the metric tensor of a
pseudo-Riemannian (below, just Riemannian) space. 
In such a way the tensorial  gravitational field was introduced.
Einstein, in this article, formulates, on the basis of some simple
model, the general energy-momentum conservation law:
\be
\pa_\nu (\sqrt{-g}\,\varTheta_\sigma^\nu)
+\f{1}{2}\sqrt{-g}\,\varTheta_{\mu\nu}\pa_\sigma g^{\mu\nu}=0.
\ee
{\it``The first three of these relations 
$(\sigma = 1, 2, 3)$ express the momentum conservation law, the
latter  
$(\sigma =4)$  that of energy conservation,''} Einstein wrote.  Here 
$\varTheta_{\mu\nu}$  stands for the energy-momentum tensor of
matter. It is necessary to note that such {\bf a law of
energy-momentum
conservation for any material system}  was introduced by Einstein
still
as a plausible physical assumption. In the same article M.~Grossman
showed that Eq.(3) is covariant under arbitrary transformations and
can be cast into the form
\be
\nabla_\nu \varTheta_\sigma^\nu =0,
\ee
here $\nabla_\nu$ is a covariant derivative with respect to the
metric $g_{\mu\nu}$. 

Einstein posed a problem to find out the gravitational equation of
the form
\be
\Gamma_{\mu\nu} =\varkappa\,\varTheta_{\mu\nu},
\ee
where $\Gamma_{\mu\nu}$ is a tensor composed of the metric and its
derivatives. 

It worth to notice that  in the part of this article which was
written by
Grossman, the possible use of the Ricci tensor,
$R_{\mu\nu}$, as $\Gamma_{\mu\nu}$ from Eq.(5),  was discussed. 

Nonetheless M.Grossman finally rejected such a proposal: {''\it It
turns out, however, that this tensor in
the special case of infinitely weak static force field 
\textbf{does
not} 
reduce itself into $\Delta\varphi$''}.

Later Einstein, following his ideas, searched for 
$\Gamma_{\mu\nu}$ as a tensor  under arbitrary  {\bf linear
transformations}. He would follow  this way till  November 1915.
At the end of June (beginning of July) 1915 Einstein,  
invited by Hilbert, 
spent a week in 
G\"ottingen, where he, as he recollected later, {\it ``gave there six
two-hour lectures''.}   
It is evident that afterwards D.~Hilbert got interested in the
problem. 

The {\bf Einstein} formulation of the problem and {\bf his}
identification of the  gravitation potential with the metric tensor 
$g_{\mu\nu}$ of a  Riemannian  space appeared the key ones for
Hilbert. That 
was sufficient for him in order to find out the gravitational field
equation
proceeding from the principle of the least action (Hilbert's Axiom~I)
and from his profound knowledge of the theory of invariants. All this
is directly seen in the paper by Hilbert~[6]. 

Below  we give an account of  Hilbert's approach to
derivation of the gravitational field equation, and also give a
critical analysis of the articles~[2,3,4] devoted to the same
question.

\section*{1. Hilbert's Approach }

Let us consider attentively Hilbert's approach~[6]. He formulates
Axiom~I:\\
\noindent

{\it The laws of  physical events are defined by the world function
$H$ the arguments of which are 
\[
g_{\mu\nu},\; g_{\mu\nu\ell}=\f{\pa g_{\mu\nu}}{\pa x^\ell},\quad
g_{\mu\nu\ell k}=\f{\pa^2 g_{\mu\nu}}{\pa x^\ell \pa x^k},
\]
\[
q_s,\; q_{s\ell}=\f{\pa q_s}{\pa x^\ell},\quad (\ell, k=1, 2, 3, 4),
\]
being the variation of the integral
\be
\int H\sqrt{g}\,d\omega,
\ee
\[
(g=|g_{\mu\nu}|,\quad d\omega=dx_1dx_2dx_3dx_4),
\]
disappears for any of 14 potentials $g_{\mu\nu},\, q_s$''}.

He writes
further: {\it``As to the world function $H$, additional axioms are
needed for its unambigious definition. If only second derivatives of
potentials $g^{\mu\nu}$ can enter the gravitation equations, then the 
function $H$ has
to have the form\footnote{In ref.[6] Hilbert used the notations
$K_{\mu\nu}$ and $K$ for the Ricci tensor and the scalar curvature.
We use for them, and also
for other quantities, modern notations. We also use in all citations
the numeration of formulas according to our text.} 
\be
H=R+L,
\ee
where $R$  is an invariant following from the Riemann  tensor 
(scalar curvature of a four-dimensional manifold):
\be
R=g^{\mu\nu}R_{\mu\nu},
\ee
\be
R_{\mu\nu}=\pa_\nu \Gamma_{\mu\alpha}^\alpha
-\pa_\alpha \Gamma_{\mu\nu}^\alpha
+\Gamma_{\mu\alpha}^\lambda \Gamma_{\lambda\nu}^\alpha
-\Gamma_{\mu\nu}^\lambda \Gamma_{\lambda\alpha}^\alpha,
\ee
and $L$  is a function of variables 
$g^{\mu\nu}, g_\ell^{\mu\nu}, q_s, q_{sk}$ only. Besides that, we
assume 
further on that $L$ does not depend on $g_\ell^{\mu\nu}$.

From variation in the  10 gravitational potentials the 10 Lagrange
differential equations follow
\be
\f{\pa \sqrt{g}R}{\pa g^{\mu\nu}}
-\pa_k \f{\pa \sqrt{g}R}{\pa g_k^{\mu\nu}}
+\pa_k \pa_\ell \f{\pa \sqrt{g}R}{\pa g_{k\ell}^{\mu\nu}}
=-\f{\pa \sqrt{g}L}{\pa g^{\mu\nu}}\;\;".
\ee
} 
{\bf It is easy to see from (8) and (9) that both in $R$ and 
$R_{\mu\nu}$ second-order derivatives of the metric enter linearly.
Second rank
tensors with such properties are }
$$
R_{\mu\nu} 
\quad \mbox{\bf and }\quad
g_{\mu\nu}R.\eqno{(10a)}
$$
{\bf All other tensors with such properties are obtained as
combinations of 
these tensors. }

This conclusion,  to some extent, was known to Einstein, and he,
mentioning tensors of the second rank, which could lead to the
gravitational equations with derivatives not higher than of the
second order, wrote in the letter to H.A.~Lorentz 19 January 1916 
[9]: 
\begin{quote}
{\it ``\ldots aside from tensors...} 
\[
R_{\mu\nu}\quad \mbox{\it and}\quad g_{\mu\nu}R
\]
{\it there are no (arbitrary substitutions for covariant) 
tensors\ldots''}
\end{quote}
For mathematician D. Hilbert   that was evident. 

Let us denote for the sake of  brevity and following to Hilbert the
left part of
the equation by the symbol 
\be
[\sqrt{g}R]_{\mu\nu}=
\f{\pa \sqrt{g}R}{\pa g^{\mu\nu}}
-\pa_k \f{\pa \sqrt{g}R}{\pa g_k^{\mu\nu}}
+\pa_k\pa_\ell\f{\pa \sqrt{g}R}{\pa g_{k\ell}^{\mu\nu}}.
\ee
Then Eq.(10) takes the form
\be
[\sqrt{g}R]_{\mu\nu}=-\f{\pa \sqrt{g}L}{\pa g^{\mu\nu}}.
\ee
Note that in  Hilbert's method of the gravitation equations
derivation one does not  need concrete specification for the
Lagrangean
function of the material system. {\bf In paper [6] D.~Hilbert
infers, in Theorem~II, the identity}:
$$
\delta_L(\sqrt{g}\,J)+\pa_\lambda(\delta x^\lambda \sqrt{g}\,J)=0\,
,\eqno{(12a)}%
$$
where $\delta_L$ is the Lie derivative; $J$ is an arbitrary function 
invariant under coordinate transformations. He uses this identity
when obtaining Eq.(48). 

{\bf Then D. Hilbert proves a very important theorem~III}: 
{\it``Let $J$ is an invariant depending only on the components of 
$g^{\mu\nu}$ and their derivatives; the variational derivatives 
of $\sqrt{g}J$ in $g^{\mu\nu}$  are designated, as earlier, as 
$[\sqrt{g}J]_{\mu\nu}$. If 
$h^{\mu\nu}$  is an arbitrary contravariant tensor then the quantity
\be
\f{1}{\sqrt{g}}[\sqrt{g}\,J]_{\mu\nu}h^{\mu\nu}
\ee
is also invariant; if one substitutes the standard tensor 
$p^{\mu\nu}$ instead of $h^{\mu\nu}$ and writes
\be
[\sqrt{g}\,J]_{\mu\nu}p^{\mu\nu}
=(i_s p^s + i_s^e p_e^s)\, ,
\ee 
where expressions 
\be
i_s=[\sqrt{g}\,J]_{\mu\nu}\pa_s g^{\mu\nu}\, ,
\ee  
\be
i_s^\ell=-2[\sqrt{g}\,J]_{\mu s} g^{\mu\ell}
\ee  
depend only on $g^{\mu\nu}$ and their derivatives,  then 
\be
i_s=\f{\pa i_s^\ell}{\pa x^\ell} \, ,
\ee
in the sense that this equation holds identically for all arguments,
i.e. 
$g^{\mu\nu}$ and their derivatives''.} 

Hilbert applies this theorem to the case 
$J=R.$ Then identity (17) assumes the form:
\be
\pa_\ell \{[\sqrt{g}\,R]_s^\ell \}
+\f{1}{2}[\sqrt{g}\,R]_{\mu\nu}\f{\pa g^{\mu\nu}}{\pa x^s}\equiv
0.
\ee
This identity is  similar to (3), hence one can write it in the form
(4)
\be
\nabla_\ell [\sqrt{g}\,R]_s^\ell \equiv 0.
\ee
We see  that the covariant derivative of the variational 
derivative 
$[\sqrt{g}\,R]_s^\ell$ is equal to zero.  Thus, on the basis of (12)
we
get 
\be
\nabla^\ell \Bigl\{\f{\pa \sqrt{g}\,L}{\pa g^{s\ell}}\Bigr\}= 0.
\ee

According to Hilbert 
the energy-momentum tensor density of the material system, 
$T_{\mu\nu}$,  is defined as follows:
\be
T_{\mu\nu}=- \f{\pa \sqrt{g}\,L}{\pa g^{\mu\nu}},
\ee 
and equality  (20) can be written down as a covariant conservation
law of the energy-momentum tensor of the material system: 
\be
\nabla_\nu T_\mu^\nu =0.
\ee

It was  Hilbert who gave for the first time the definition (21) of
the
energy-momentum tensor of the material system and showed that this
tensor satisfies Eq. (22); by that he gave a basis of  the Einstein's
assumption
from Ref.~[8]. So, D.~Hilbert {\bf found the gravitational field
equation}\footnote{Original paper [6] by Hilbert corresponds to the
system of units  where $\varkappa=1$. {\sl Authors}}
\be
[\sqrt{g}\,R]_{\mu\nu}=-\varkappa\,T_{\mu\nu},
\ee
from which {\bf the law of covariant conservation of energy-momentum
(22) 
follows exactly}.

Multiplying both parts of Eq.~(23) by $g^{\mu\nu}$ 
and summing up in indices 
$\mu$ and $\nu$ we get
\be 
g^{\mu\nu}[\sqrt{g}\,R]_{\mu\nu}=-\varkappa\, T \,.
\ee
In the l.h.s. of Eq.(24) an invariant is formed, which contains
second
derivatives linearly. But there exists only one such invariant, $R$. 
One gets thereof the equation
\be
\sqrt{g}\,\beta R=-\varkappa\, T \, ,
\ee
where $\beta$ is an arbitrary constant. 

Summarizing one can say that the gravitational field equations were
found 
{\bf by D.~Hilbert and, by that, the problem, posed by A.~Einstein in 
1913 was resolved. Eq.~(23) are identical with Eq.(1). They differ
only in the form. Below we will see that, according to Hilbert,
Eqs.~(23)
are easily transformed into (1).} Hilbert, both in the proofs and in
paper~[6], wrote: 
\textit{``\textbf{In the following I want} \ldots \textbf{to
establish} \ldots
\textbf{a new system of fundamental equations of physics}''.}
And further: \textit{``\ldots \textbf{my fundamental equations}''\,,
``\ldots
\textbf{my theory}\ldots''.}
D.~Hilbert could not write so, if he did not considered 
{\bf himself} the author  of the ``fundamental equations of
physics''.

The tensor density $[\sqrt{g}\,R]_{\mu\nu}$ in Eq.~(23) 
contains by construction the second order derivatives linearly, so,
on the basis of
(10a) this energy density has the form 
\be
[\sqrt{g}\,R]_{\mu\nu}=\sqrt{g}(R_{\mu\nu}+\alpha g_{\mu\nu}R)\,
.
\ee
Expression (26) was quite evident for Hilbert. Maybe for the authors
of~[2,3,4] it is difficult to understand that, but this is their
personal affair. For the l.h.s. of Eq.~(24) one obtains, on the
basis of (26),
\be
g^{\mu\nu}[\sqrt{g}\,R]_{\mu\nu}=\sqrt{g}(4\alpha+1)R,
\ee
which is in complete correspondence with (25). Namely about these
general reasonings Hilbert wrote: 
{\it``\ldots what is  clear without calculations if to take into
account that $R$ is the only invariant and 
$R_{\mu\nu}$ is the only (besides
$g_{\mu\nu}$) second-order tensor, which can be constructed from  
$ g_{\mu\nu}$ and its first  and second  derivatives 
$g_k^{\mu\nu},  g_{k\ell}^{\mu\nu}$''.}

Authors of  paper [2] (see also [3]) write in this connection
{\it ``This argument is, however, untenable, because there are many
other tensors of second rank and many other invariants that can be
constructed from the Riemann tensor''.}

{\bf This statement of the authors of [2] has no relation to the
exact Hilbert's argument because the authors of papers [2,3]
overlooked the main thing: one argued on the construction of the
gravitational equations containing  derivatives of 
$g^{\mu\nu}$ of order not higher than two.}
Hilbert specially wrote about that in his paper~[6]:
{\it ``If only second order derivatives of the potentials 
$g^{\mu\nu}$ can enter the gravitational equations, then the function
$H$
has to have the form}
\[  
H=R+L \,".
\]
Therefore D.~Hilbert was {\bf absolutely right}  that in this case
only  $R$ and {\bf two tensors},  
$R_{\mu\nu}$ and 
$g_{\mu\nu}R$, 
contain linearly second derivatives of the gravitational potential 
$g^{\mu\nu}$. All other tensors with such properties are linear
combinations of these tensors.

Likewise the author of the paper [4] is wrong when he writes: 
{$\ll$\it \ldots va\-ri\-a\-ti\-o\-nal derivation of the equations
is
absent, and
the
right form of the equations (with the ``half'' term) is motivated
(not quite correctly) by the uniqueness of the Ricci tensor and
the scalar curvature as generally covariant quantities, depending
only on
$g^{\mu\nu}$'s and their first and second derivatives$\gg$}.

It is  astonishing indeed when the author of the paper [4] writes
about
Hilbert's paper: {\it ``...variational derivation... is absent''}. 
He probably {\bf forgot} a well known circumstance that the Lagrange
equations, which were presented by Hilbert, are a consequence of the
least action principle (Axiom~I of Hilbert). Thus, {\bf the
variational
derivation of the gravitational field equation takes place in
Hilbert's paper~[6]}. 

How the authors of [2,3,4] make up their mind to analyze and to judge
D.~Hilbert's papers~[6] if they  do not understand the essence of his
exact mathematical arrguments? The authors of papers [2,3] write
further: {\it ``Even if one requires the tensors and invariants to be
linear in the Riemann tensor, the crucial coefficient of the trace
term remains undetermined by such an argument''}. This is again {\bf
wrong}, it is easily determined. Hilbert proved the identity (19)
\be
\nabla_\sigma[\sqrt{g}\,R]_\mu^\sigma\equiv 0 \, .
\ee
With account of Eq.~(26) and with use of the local frame where
Christoffel's symbols are zero, the identity (28) takes a simple
form: 
\be
\pa_\sigma(R_\mu^\sigma +\alpha\delta_\mu^\sigma\,R)\equiv 0 \,
.
\ee
From (8) and (9) one finds 
\be
\pa_\mu R=K_\mu,\quad \pa_\sigma R_\mu^\sigma =\f{1}{2}K_\mu \,
,
\ee 
where 
\be
K_\mu =g^{\nu\sigma}g^{\lambda\rho}\pa_\sigma \pa_\nu \pa_\mu 
g_{\lambda\rho} 
-g^{\nu\sigma}g^{\alpha\lambda}\pa_\sigma \pa_\alpha \pa_\mu 
g_{\lambda\nu}\, .
\ee 
Making use of these expressions we get
\[
\pa_\sigma(R_\mu^\sigma +\alpha\delta_\mu^\sigma\,R)
=\left(\f{1}{2}+\alpha\right)K_\mu\equiv 0\, .
\]
We have thereof 
\be
\alpha=-\f{1}{2},
\ee
and hence,
\be
[\sqrt{g}\,R]_{\mu\nu}=\sqrt{g}\left(R_{\mu\nu}-\f{1}{2}g_{\mu\nu}R
\right)\,
,
\ee
i.e. 
\be
\sqrt{g}\left(R_{\mu\nu}-\f{1}{2}g_{\mu\nu}R\right)=-\varkappa\,
T_{\mu\nu}\,
.
\ee

Thus {\it ``the critical coefficient''},  that is a concern of the
authors
of [2,3],  is obtained in Hilbert's approach in a trivial way by
taking
derivatives fairly accessible to a first-year student of a
university. It is also clear that the trace term 
$\f{1}{2}g_{\mu\nu}R$  does not arise as a result of some arbitrary
``introduction'' into the field equations formulated by Hilbert; it
{\bf is organically contained} there. 

Later, in 1921, in paper [10], A.~Einstein would construct the 
geometrical part  of the gravitational equations making use of the
tensor 
\[
R_{\mu\nu} + ag_{\mu\nu}R,
\]
i.e. in the same way as it was done earlier by Hilbert at the
transformation of gravitation equations (12) to the form (34).
Creative endeavour of the authors of~[2,3] is crowned with the
following thoughtful conclusion: {\it ``Taken together, this sequence
suggests that knowledge of Einstein's result may have been crucial to
Hilbert's introduction of the trace term into his field equations''}.

How can one arrive to such an idea after reading Hilbert's paper? Let
us remind to the authors of~[2] that, {\textbf {in Hilbert's
formalism, one
does not need to introduce anything. As soon as one  wrote the world 
function $H$ in the form}}
{
\mathversion{bold}
\[ 
H =R+L,
\]
}
{\textbf{and established Theorem III, the rest was just a matter of
calculational techniques, and nothing more.}}

{\textbf{Thus, the analysis that we have undertaken on the judgements
of
the authors of~[2] shows that all their reproofs to Hilbert are
either wrong or do not concern him. So all their arguments in favour
of the point of view \textit{``that radically differs''} from the 
standard one are inconclusive.} 

Hilbert certainly obtained, before publication of Einstein's paper
with the trace
term, the equality (33). Taking use of (19) and (33)  we find
\be
\nabla_\nu\left(R_\mu^\nu -\f{1}{2}\delta_\mu^\nu R\right)\equiv
0.
\ee
{\bf But this  is the Bianchi identity.}

Poor knowledge of Hilbert's paper can be met not only in Refs.[2].
For
instance, A.~Pais in the book~[11], \S 15.3, wrote: ``{\it Evidently
Hilbert did not know the Bianchi identities either!''} and
further: {``\it I repeat one last time that neither Hilbert nor
Einstein was aware of the Bianchi identities in that crucial November
1915''}. {\it ``Interesting enough, in 1917 the
experts
were not aware that Weyl's derivation of Eq.~15.4} (The identity in
question. ---{\sl Authors}) {\it by variational techniques was a
brand new
method for obtaining a long-known result''}.
A.~Pais was right in that A.Einstein did not
know the Bianchi identity in that crucial November 1915.   {\bf All
the
rest in~[11],
concerning Hilbert, is wrong. The matter is that Hilbert did not know
the Bianchi identity, indeed. He just himself obtained it.}
D.~Hilbert
proved with variational method the general identity (see Theorem~III
by Hilbert), from which, putting 
\( J=R \), he obtained also the Bianchi identity. 

{\bf Thus it was not Weyl in 1917 but Hilbert in 1915 who obtained
the
Bianchi identity with variational method.} A.~Pais wrote in
\S~15.3 {``\it In November 1915, neither Hilbert nor Einstein was
aware of this royal road to the conservation laws. Hilbert had come
close''}. 

The authors of~[3] write similarly: {\it ``\ldots... Hilbert did
not discover  royal road to the formulation of the field equations of
general relativity. In fact, he did not formulate these equations at
all\ldots''}. 

All this is wrong. Namely Hilbert found the shortest and general way
to formulate the gravitational equations. He found the Lagrange
function of the gravitational field, $R$, with help of which the
gravitational equations are obtained automatically via the principle
of the least action. {\bf One obtains them namely in such a way when
giving an account of Einstein's General Relativity.} It is a pity
that
A.~Pais seems to look through the Hilbert paper superficially; the
same is
true for the authors of Refs.~[2,3]. 

Later, in 1924, D.~Hilbert wrote~[12]: {``\it In order to define the
expression \( [\sqrt{g}\,R]_{\mu\nu} \) one chooses first the frame
in such a way that all 
\( {g}_s^{\mu\nu}, \)  taken in the world point, disappear. We find
thereof 
\be
[\sqrt{g}\,R]_{\mu\nu}=\sqrt{g}\Bigl[R_{\mu\nu}-\f{1}{2}g_{\mu\nu}R
\Bigr]\,
".
\ee
}
Authors of [2] write, concerning this: {\it ``To summarize: Initially
Hilbert did not give the explicit form of the field equations; then,
after Einstein had published his field equations, Hilbert claimed
that no calculation is necessary; finally, he conseded that one
is.''}. 

This statement is a creation of the mind of the authors of~[2]. No
solid reasons exist  that Hilbert did not obtained, himself,  the
explicit form of
the field equations. One obtains them  in an elementary way from
Eqs.(23) and
expression (26) with the use of the identity (28). {\bf If one can
seriously assume that Hilbert was not able to obtain (33) from (28)?}
Hilbert's addition made in 1924 does not mean a ``recognition that
calculation is necessary''. He introduced it just to remind a simple
method to find a tensor. This did not discard his exact argument
(``...clear without calculation'') at all. 

The authors of [2,3] claim, referring to the Proofs, that Hilbert had
the gravitation equation only in the form (23). Equation (23)
contains the derivatives 
\be
\f{\pa\sqrt{g}\,R}{\pa g^{\mu\nu}},\quad \f{\pa\sqrt{g}\,R}{\pa
g_k^{\mu\nu}},\quad
\f{\pa\sqrt{g}\,R}{\pa g_{k\ell}^{\mu\nu}}.
\ee
It is impossible  to imagine a physicist-theorist or mathematician
who
would not calculate these derivatives and obtain explicitly the
differential equations containing only derivatives 
\( g_k^{\mu\nu}, g_{k\ell}^{\mu\nu} \). 
As we have seen, it was not nesessary, for Hilbert, to calculate
them, because he managed to identify the structure of the expression 
\( [\sqrt{g}\,R]_{\mu\nu} \) from the general and rigorous
mathematical statements, due to which the calculation of the {\it
``critial coefficient''} became trivial.

\textbf{That is why the conclusion of the authors of papers [2,3,4],
that Hilbert did not obtain the \textit{``explicit form of the
gravitational
field equations''}  cannot be true.} It contradicts also, as we will
see further, to the correspondence between Einstein  and Hilbert,
from which everything becomes absolutely clear, and no additional
arguments are needed. There does not exist more decisive argument
than the  evidence of Einstein himself. {\bf But precisely this most
important evidence of Einstein was left afield by the authors of
[2,3]}, who put into the center of their analysis unpublished and
mutialted materials  of Hilbert. 

The Einstein evidence in his letter to Hilbert of 18 November 1915
excludes unambiguously any false conjectures about Hilbert's
paper~[6]. Thus, the \textbf{``\textit{archive finding''}} of the
authors of
[2],
as a matter of principle, cannot shatter the evidence of Einstein
himself.
One could stop here the further discussion of the question. But
the authors of [2,3,4] alongside their arguments make erroneous
conclusions about Hilbert's paper~[6]. So we would like  to
specially concentrate on this. 

Even if one does not follow the general statements of Hilbert, it us
still possible, making use of definition (11), to execute simple
differentiation and to express the tensor density 
\( [\sqrt{g}\,R]_{\mu\nu} \) in terms of the Ricci tensor density and
scalar density 
\( \sqrt{g}\,R \).
The first term in (11) can be written in the form
\be
\f{\pa\sqrt{g}\,R}{\pa g^{\mu\nu}}
=\sqrt{g}
\left(R_{\mu\nu}+\f{1}{\s g}\;\f{\pa\sqrt{g}}{\pa g^{\mu\nu}}R\right)
+\sqrt{g}\,g^{\alpha\beta}\f{\pa R_{\alpha\beta}}{\pa g^{\mu\nu}}
\,,
\ee
Because of 
\be
\f{\pa\sqrt{g}}{\pa g^{\mu\nu}}
=-\f{1}{2}\sqrt{g}\,g_{\mu\nu} \, ,
\ee
we get 
\be
\f{\pa\sqrt{g}\,R}{\pa g^{\mu\nu}}
=\sqrt{g}\left(R_{\mu\nu}-\f{1}{2}g_{\mu\nu}\,R\right)
+\sqrt{g}\,g^{\alpha\beta}\f{\pa R_{\alpha\beta}}{\pa g^{\mu\nu}} \,
.
\ee
We have on the basis of (11) and (40):
\[
[\sqrt{g}\,R]_{\mu\nu}
=\sqrt{g}\left(R_{\mu\nu}-\f{1}{2}g_{\mu\nu}\,R\right)+
\Biggl\{\sqrt{g}\,g^{\alpha\beta}\f{\pa R_{\alpha\beta}}{\pa
g^{\mu\nu}}
-\pa_k \f{\pa \sqrt{g}\,R}{\pa g_k^{\mu\nu}}
+\pa_k \pa_\ell\f{\pa \sqrt{g}\,R}{\pa
g_{k\ell}^{\mu\nu}}\Biggr\}.
\]
It is easy to see that terms in figure parenthesis dissapear
identically. The most simple way is to use the local Riemannian
frame  where Christoffel symbols are zero. In such a simple, but not
very elegant, way we arrive again to the expression
\[
[\sqrt{g}\,R]_{\mu\nu}=\sqrt{g}\Bigl(R_{\mu\nu}-\f{1}{2}g_{\mu\nu}R
\Bigr).
\]
The authors of paper [3] wrote: ``{\it In both the Proofs and the
published version of  paper~[6], Hilbert erroneously claimed
that one can consider the last four equations} (i.e. electromagnetic
field equations. ---{\sl Authors}) {\it as a consequence  of the
4~identities that must hold, according to his Theorem~I, between the
14 differential equations\ldots''}.

Things, however, are not such as the authors of [3] suppose.
Theorems~I and
II are formulated for $J$, an invariant under arbitrary
transformations of the four world parameters. According to these
theorems, there exist {\bf four identities} for any invariant.
Hilbert, in his paper, considers two invariants, $R$ and $L$. The
general invariant $H$ is composed of these two invariants:
\[
H=R+L \,.
\]

The gravitation equations, in Hilbert's notations, have the form:
\[
[\sqrt{g}\,R]_{\mu\nu}=-\varkappa\,T_{\mu\nu}\, ,
\]
Hilbert chooses the invariant  \( L \) as a function of the variables
\( g^{\mu\nu}, q_\sigma, \pa_\nu q_\sigma \) and so he obtains the
generalized Maxwell equations
\be
[\sqrt{g}\,L]^\nu =0,
\ee
where
\be
[\sqrt{g}\,L]^\nu =\f{\pa \sqrt{g}\,L}{\pa q_\nu}
-\pa_\mu \left(\f{\pa \sqrt{g}\,L}{\pa (\pa_\mu q_\nu)} \right).
\ee

Then, on the basis of Theorem II,  Hilbert obtains that the Lagrange
function $L$ depends on the derivatives of the potential 
\( q_\nu \) only via the combination 
\( F_{\mu\nu} \), i.e.
\be
L(F_{\mu\nu})\, ,
\ee
where 
\be
F_{\mu\nu}=\pa_\mu q_\nu - \pa_\nu q_\mu \, . 
\ee
On this basic Hilbert chooses the Lagrangean in the form 
\be
L=\alpha Q+f(q) \, ,
\ee
where
\be
Q=F_{\mu\nu}F_{\lambda\sigma}g^{\mu\sigma}g^{\nu\lambda},\quad
q=q_\mu q_\nu g^{\mu\nu},
\ee
here  \( \alpha \) is a constant. 

Hilbert then remarks that the equations of electrodynamics 
{\it ``can be considered as a consequence of the equations of
gravitation''}.

According to Theorem II the {\bf four identities} take place for the
invariant~\( L \): 
\be
\nabla_\mu T_\nu^\mu=F_{\mu\nu}
[\sqrt{g}\,L]^\mu +q_\nu \pa_\mu [\sqrt{g}\,L]^\mu .
\ee
It follows from identity (47) that, if the equations of motion of a
material system (41) hold, then the covariant conservation law takes
place:
\[
\nabla_\mu T_\nu^\mu=0. 
\]
If one makes use of the gravitation equations (34) for identity (47)
then Hilbert's equations result: 
\be
F_{\mu\nu}[\sqrt{g}\,L]^\mu +q_\nu \pa_\mu [\sqrt{g}\,L]^\mu
=0,
\ee
which were denoted in his paper [6] under the number (27). 

Equations (48) have to be {\bf compatible with the equations, which
follow
from the principle of the least action with the same Lagrangean
{\mathversion{bold}\( L \)}.}
It is only possible in the case, when the {\bf ``generalized Maxwell
equations''} hold:
\be 
[\sqrt{g}\,L]^\nu=0.
\ee

Therefore, the author of paper [4] is completely wrong, considering
that {\it ``in the case of gauge-noninvariant Mie's theory with a
Lagrangean of the kind (45) one has in general use not the
generalized Maxwell equations (49), but rather equations (48).''}

This statement contradicts  the principle of the least-action, i.e.
Hilbert's Axiom~I. 
{\bf So, the four identities (47) due to Theorem~II and equations of
gravitation (34) lead to the four equations (48) which are
compatible
with the generalized Maxwell equations, obtained on the basis of
Hilbert's Axiom~I}. This is what Hilbert emphasized in paper~[6].
In this relation he pointed out: \textit{``\ldots from the
gravitation
equations (10) there really follow 4 mutually independent linear
combinations (48)  \textbf{of equations of the electrodynamics
(41)}} (emphasized by the authors) \textit{altogether with their
first
derivatives''.}

One has to specially stress that Hilbert writes about {\it ``linear
combinations of the equations of electrodynamics (41)''}, but not the
expressions (42). Namely here the authors of [3,4] admit a confusion. 

Let us note that in the particular case, when
\be
L=\alpha Q,
\ee
the second  term in Eq.(48) dissapears identically and we come to the
equations
\[
F_{\mu\nu}[\sqrt{g}\,L]^\mu =0.
\]
It follows therefore that if the determinant  \( |F_{\mu\nu}| \)  is
not zero, the Maxwell equations take place
\[ 
[\sqrt{g}\,L]^\mu=0,
\]
which are in full agreement with the principle of the least action
(Hilbert's Axiom~I). {\bf In such a way the Maxwell equation are the
consequence of the gravitation equation (34) and four identities
(47).}  All this follows from Hilbert's article if one reads it
attentively. Afterwards Einstein together with Infeld and Hoffmann in
[13], and also Fock in [14] would obtain the equation of motion of
a material system from the gravitation equations.

One notices quite often that Hilbert obtained the gravitational field
equation ``{\it \ldots  not for an arbitrary material system, but
especially  basing on Mie's theory''}~[15]. That is not quite right.
Method which Hilbert used is general and no limitations are implied
on the form of the function~$L$. 

The circumstance that the gravitation equations imply {\bf four
equations} for the material system, looked attractive for Hilbert and
he applied his general equations to Mie's theory. Such a unification
of gravitation and Mie's theory was not fruitful, but Hilbert's
general
method for obtaining the gravitation equations proved to be very
far-reaching.

Now a few words about auxialiary noncovariant equations.

To solve a problem it is always necessary to have a complete system
of equations. There are only ten equations of general relativity. One
still needs to add four equations, which cannot be chosen generally
covariant. These auxiliary conditions are called coordinate
conditions, and can be of various kinds. Hilbert meant namely this
when
he wrote (see Proofs in [7]): {\it ``As our mathematical Theorem
shows us,
the previous Axioms~I and II can give only 10 mutually independent
equations for 14 potentials. On the other hand, due to general
invariance, more than 10 essentially independent equations for 14
potentials, 
\( g_{\mu\nu}, q_s \), are  impossible, and, as we wish to hold on
Cauchy's theory for differential equations and to give to basic
equations of physics a definite character, an addition to (4) and (5)
of 
auxiliary non-invariant equations is inevitable.''}

This is a mathematical requirement and it is necessary for a theory.
Hilbert tried to obtain these additional equations in the framework
of the very theory, but he failed to do this and did not include that
into the published article.

So the basic system of the 10 equations of general relativity is
generally covariant, but the complete system of equations which is
necessary to solve problems is not generally covariant because four
equations expressing coordinate restrictions cannot be tensorial;
they 
are  not generally covariant. The solution to a complete system
of the gravitational field equations can be always written in any
admissible coordinate system. Namely here a notion of the chart atlas
arises. 

{\bf That is why the statement of the authors of [2,3,4] that
Hilbert's theory is not generally covariant, in contrast with
Einstein's theory, is wrong. The complete system of equations both of
Hilbert and Einstein is not generally covariant. }

The only difference was in that Hilbert tried to uniquely construct
these non-covariant equations in the framework of the very theory.
This appeared impossible. The equations defining the choice of frame
became quite arbitrary but not tensorial. 

In this relation J.~Synge~[16] writes:``{\it One can find in the
papers on general relativity a number of various coordinate
conditions, pursuing every time special aims. In order to approach
the problem in a unified way let us write down the coordinate
conditions in the form
$$
C_i=0,\quad (i=0, 1, 2, 3).
$$
The metric tensor $g_{ij}$ must satisfy these equations (perhaps
differential). Certainly, they \textbf{cannot be} tensorial, because
they are satisfied only at a special choice of coordinates.
}''.
What is the material on which the authors of  paper [2]
made their conclusions? In the so-called proofs of D.~Hilbert's
paper,  they proceeded from, the invariants $H$ and $K$ are used but
there is no
their definition. D.~Hilbert writes in the Proofs: {\it``I would like
to construct below a new system of basic equations of physics,
following  the  axiomatic method and proceeding, essentially, from
the
three axioms''.}

Evidently, Hilbert had to define the invariants $H$ and $K$ in order
to do that. It is impossible to imagine that Hilbert, having posed
such an aim, did not define these fundamental quantities. {\bf But
this
means that the parts absent from  the Proofs are very essential and
contain an important information. Valid conclusions cannot be  made
without account of this key infromation.}

However the authors of [2] neglected  this important circumstance
and were in a hurry to conclude that Hilbert did not derive the
gravitation equations in the form
\[
\sqrt{g}\left(R_{\mu\nu}-\f{1}{2}g_{\mu\nu}R\right)
=-\varkappa T_{\mu\nu} \, .
\]
They presented this conclusion to the wide scientific community in a
popular and well-known journal ``Science''~[2]. For all that the  
authors of [2] did not inform the readers that so-called Proofs are
mutilated. Only later, in~[3], they mentioned that. The authors
of [2] claim that the Proofs allowed them to base their point of
view ``that radically differs from the standard'' one. How could it
be done on the basis of a preliminary and mutilated material? 

Here is one more method of ``analysis'' used by the authors of~[3]:
\textit{``Remarkably, in characterizing his system of equations,
Hilbert
deleted the word ``\textbf{neu}'',  a clear indication that he had
meanwhile
seen Einstein's paper and recognized that the equations implied by
his own variational principle are formally equivalent to those which
Einstein had explicitly written down (because of where the trace
term occurs), if Hilbert's stress-energy tensor is substituted for
the unspecified one on the right-hand side of Einstein field
equations.''} 

But all cited above loses sense because actually their
\textbf{``\textit{clear
indication}''} disappears as D.~Hilbert in the published article~[6]
wrote quite clearly: {\it ``I would like to construct below\ldots a
new system of fundamental equations''}. 

It is extremely tactless to produce conclusions on Hilbert's
ideas on the basis of his marginal remarks in preliminary unpublished
materials. The system of gravitational equations obtained by Hilbert
is really the {\bf new} one. He obtained it without knowledge that
A.~Einstein came to the same gravitational equations. That is what 
A.~Einstein wrote to D.~Hilbert about  in the letter of 18 November
1915 (see Section~3). Strange is the way, chosen by the authors of
[3], to base their ``radically different'' point of view. Many-page
composition [3] abounds in both similar doubtful arguments and
erroneous statements. Such an approach to the study of most important
physics papers can be hardly considered as a professional, based on a
profound analysis of the material.

In conclusion of this section let us note, that Hilbert's papers
under general title {\bf \textsf{``Grundlagen der Physik''}} are very
important and
instructive. It would be very good if theoreticians, who deal with
similar
problems, knew them.

Thus, for instance, an article [17] was published in ``Uspekhi''.
Should the authors of this paper read Hilbert's paper [18],
published in 1917, they would see that the critical coordinate
velocity ${\scriptstyle V_c}$, which they calculated approximately,
is equal in fact to 
\[
{\scriptstyle V_c}=\f{1}{\sqrt{3}}\left(\f{r-\alpha}{r}\right),
\quad \alpha =r_g=2GM.
\]
Namely at this velocity the acceleration is equal to zero. Velocity 
\( {\scriptstyle V_c} \) depends on the radius, while the
corresponding
proper velocity, 
\( v \), does not depend on 
\( r \) and
\[ 
v=\f{1}{\sqrt{3}} \, .
\] 
In order to obtain the critical coordinate velocity  \( {\scriptstyle
V_c} \) in the first order in 
\( G \) one needs to keep in acceleration terms of the 
second order in  \( G \). 
Gravitational field does not exert an action on a body, if the latter
moves with velocity 
\( {\scriptstyle V_c} \), under the action of some external force.  

In paper [18] D.~Hilbert obtains the equation
\[
\f{d^2 r}{dt^2}-\f{3 \alpha}{2r(r-\alpha)}
\left(\f{dr}{dt}\right)^2+\f{\alpha (r-\alpha)}{2r^3}=0
\]
and adduces its integral:
\[
\left(\f{dr}{dt}\right)^2=
\left(\f{r-\alpha}{r}\right)^2+
A\left(\f{r-\alpha}{r}\right)^3,
\]
where \( A \) is a constant; for the light  \( A=0 \). 

One obtains thereof the formula (20) for the  velocity from the paper
[17]
\[ 
\left(\f{dr}{dt}\right)^2=
\f{1}{3}\left(1-\f{r_g}{r}\right)^2
\left(1+\f{2r_g}{r}\right)\, ,
\]
which differs from the critical velocity  
\( {\scriptstyle V_c} \). 
At this velocity the acceleration is not zero. 

D.~Hilbert writes further:
{\it `` According to this equation the acceleration is negative or
positive, i.e. gravitation attracts or repulses dependent on if the
absolute value of the velocity obeys to inequality}
\[
\left|\f{dr}{dt}\right|<\f{1}{\sqrt{3}}\left(\f{r-\alpha}{r}\right),
\]
{\it or inequality} 
\[
\left|\f{dr}{dt}\right|>\f{1}{\sqrt{3}}\left(\f{r-\alpha}{r}\right)".
\]
For the light Hilbert finds
\[
\left|\f{dr}{dt}\right|=\f{r-\alpha}{r},\;
\]
and further
he notes:
{\it ``The light propagating rectilinearly towards the center
experiences always a repulsion according to the latter inequalities;
its speed increases from 0 at 
\( r=\alpha \) to 
\( 1 \) at \( r=\infty \)''.}

Let us note that the local speed of light is equal to 1 (in units of
\( c \)).
It is also necessary to note that the  velocity  \( {\scriptstyle
V_c} \) is
not
a solution of the initial equation.

One more remark. The authors of [17] write: {$\ll$\it Maybe this is
the
reason why sometimes in the literature the proper time is called 
``genuine'', or ``physical''. A lightminded person would think that
any other time (the coordinate time) is not physical, and thus should
not be considered$\gg$}. 

And further: {\it ``As a result some many specialists on general
relativity consider coordinate-dependent quantities as
nonphysical, so to say {\it ``second-quality''}
quantities. However the coordinate time is even more important for
some problems than the proper time $\tau$''}.

So, as the authors of [17] notice: {\it $\ll$... to speak about the
proper time as a ``genuine'' or ``physical'' in contrast with the
coordinate velocity is  not logical$\gg$.
} In vain the authors of [17] 
think that specialists in general relativity do not understand
significance of coordinate quantities. All the description in general
relativity proceeds in terms of coordinate quantities. One cannot
avoid them in principle. This is well known for a long time.

As an example of the physical quantity let us take the  proper time,
which
differs from the coordinate one in that it does not depend on the
choice of the coordinate time. As one sees there is a difference, and
it is quite essential. Another  example is the coordinate velocity of
light
\[
{\scriptstyle V}=
\f{\sqrt{g_{00}}}{1-\ds\f{g_{0i}e^i}{\sqrt{g_{00}}}}\,
,
\]
here \( i=1, 2, 3;\;e^i \) is a unit vector in the three-dimensional
Riemannian space.

The coordinate velocity 
 \( {\scriptstyle V} \) is, certainly , measurable but depends on the
 choice of coordinates and can have an arbitrary value:
\[ 
0<{\scriptstyle V}<\infty \, ,
\]
while the physical speed of light is equal exactly to 
1 (\( c \)). As one can see there is also a difference, and also very
essential. 

Therefore there is nothing ``non-logical'' in the use of notions of
physical and coordinate velocities, contrary to the authors of [17].

\section*{\Large 2. A. Einstein's approach}


Einstein wrote in 1913~[8]:
{\it ``The theory stated in the following arose from the conviction
that proportionality between the inertial and gravitational masses of
bodies is an exact, real law of Nature, which must find its
expression already  in the basis of theoretical physics. Already  in
some earlier works I tried to express this conviction reducing the
{\bf\textit{gravitational}} mass to the {\bf\textit{inertial}} one;
this aspiration led me to the
hypothesis that the gravity field (homogeneous in an infinitesimally
small volume) can be physically substituted by an accelerated frame
''}. 

Namely this path led Einstein to the conviction that in general case
the gravitational field is characterized by the ten space-time
functions (metric coefficient of the Riemann space) \( g_{\mu\nu} \)
\be
ds^2=g_{\mu\nu}(x)dx^\mu dx^\nu.
\ee

He further published a series of papers  about which he wrote
later in the paper [19]:
{\it $\ll$ My efforts in recent years  were directed toward basing a
general
theory of relativity, also for nonuniform motion, upon the
supposition of relativity. I believed indeed to have found the only
law of gravitation that complies with a reasonably formulated
postulate of general relativity; and I tried to demonstrate the truth
of precisely this solution in a paper\footnote{Equations of this
paper are quoted in the following with the additional note ``l.c.''
in order to keep them distinct from those in the present paper.} [8]
that appeared last year in
the ``Sitzungsberichte''. 

Renewed criticism showed to me that this truth is absolutely
impossible to show in the manner suggested. That this seemed to be
the case was based upon a misjudgment. The postulate of relativity
---
as far as I demanded it there --- is always satisfied if the
Hamiltonian principle is chosen as a basis. But in reality, it
provides no tool to establish the Hamiltonian function $H$ of the
gravitational field. Indeed, equation (77)l.c. which limits the
choise of $H$ says only that $H$ has to be an invariant toward linear
transformations, a demand that has nothing to do with the relativity
of accelerations. Furthermore, the choice determine by equations
(78)l.c. does not determine equation (77) in any way. 

For these reasons I lost trust in the field equations I had derived,
and instead looked for a way to limit the possibilities in a natural
manner. In this pursuit I arrived at the demand of general
covariance, a demand from which I parted, though with a heavy heart,
three years ago when I worked together with my friend Grossmann. As a
matter of fact, we were then quite close to that solution of the
problem, which will be given in the following. 

Just as the special theory of relativity is based upon postulate that
all equations have to be covariant relative to linear orthogonal
transformations, so the theory developed here rests upon the
\textbf{postulate of the covariance  of all systems of equations
relative to
transformations with the substitution determinant~1.}

Nobody who really grasped it can escape from its charm, because it
signifies a real triumph of the general differential calculus as
founded by Gauss, Riemann, Christoffel,  Ricci, and Levi--Civita.
$\gg$}

Einstein chose the gravitation equation in the coordinate system 
$\sqrt{-g}=1$ in the form\footnote{In this Section we use Einstein's
notations. ({\sl Authors)}}
\be
\pa_\alpha\Gamma_{\mu\nu}^\alpha+\Gamma_{\mu\beta}^\alpha\Gamma_{\nu\
alpha}^\beta
=-\varkappa\, T_{\mu\nu}\, ,
\ee
where
\[
\Gamma_{\mu\nu}^\alpha=-\f{1}{2}
g^{\alpha\sigma}(\pa_\mu g_{\nu\sigma}+\pa_\nu g_{\mu\sigma}
-\pa_\sigma g_{\mu\nu}) \, ,
\]
being \( T_{\mu\nu} \) the energy-momentum tensor for a material
system. The l.h.s. of Eq.(52) is obtained from the Ricci tensor at
the condition
\( \sqrt{-g}=1 \). 

Einstein finds the Lagrange function for the gravitational field
\be
L=g^{\sigma\tau}\Gamma_{\sigma\beta}^\alpha\Gamma_{\tau\alpha}^\beta
\, .
\ee
If one takes into account the relation
\be
2\Gamma_{\sigma\beta}^\alpha\delta(g^{\sigma\tau}\Gamma_{\tau\alpha}^
\beta)
=\Gamma_{\sigma\beta}^\alpha\delta g_\alpha^{\sigma\beta}\, ,
\ee
then it is easy to obtain:
\be
\delta L=-\Gamma_{\sigma\beta}^\alpha\Gamma_{\tau\alpha}^\beta
\delta g^{\sigma\tau}
+\Gamma_{\sigma\beta}^\alpha \delta g_\alpha^{\sigma\beta} \, .
\ee
We have thereof 
\be
\f{\pa L}{\pa
g^{\mu\nu}}=-\Gamma_{\mu\beta}^\alpha\Gamma_{\nu\alpha}^\beta,\quad
\f{\pa L}{\pa g_\alpha^{\mu\nu}}=\Gamma_{\mu\nu}^\alpha \, .
\ee
With help of these formula the gravitation equation (52) can be cast
into  the form
\be
\pa_\alpha \left(\f{\pa L}{\pa g_\alpha^{\mu\nu}}\right)
-\f{\pa L}{\pa g^{\mu\nu}}=-\varkappa\, T_{\mu\nu} \, .
\ee
Multiplying (57) by \( g_\sigma^{\mu\nu} \) and summing up in indices 
\( \mu \) and \( \nu \), Einstein obtains
\be
\pa_\lambda t_\sigma^\lambda=
\f{1}{2}T_{\mu\nu}\pa_\sigma g^{\mu\nu} \, ,
\ee
where the quantity 
\be
t_\sigma^\lambda=
\f{1}{2\varkappa}\left(\delta_\sigma^\lambda L
-g_\sigma^{\mu\nu}\f{\pa L}{\pa g_\lambda^{\mu\nu}}\right) \, ,
\ee
characterizes the gravitational field. Taking into account the
equality:
\[
\Gamma_{\mu\nu}^\lambda \pa_\sigma g^{\mu\nu}
=2g^{\alpha\mu}\Gamma_{\alpha\sigma}^\nu\Gamma_{\mu\nu}^\lambda \, ,
\]
one finds:
\be
t_\sigma^\lambda=
\f{1}{\varkappa}\left(\f{1}{2}\delta_\sigma^\lambda g^{\mu\nu}
\Gamma_{\mu\beta}^\alpha\Gamma_{\nu\alpha}^\beta 
-g^{\alpha\mu}\Gamma_{\alpha\sigma}^\nu\Gamma_{\mu\nu}^\lambda\right)
\,
.
\ee      
All further calculations are made in the reference frame, where 
\( \sqrt{-g}=1 \). Einstein writes down the basic equations of
gravitation (52) in the form
\be
\pa_\alpha(g^{\nu\lambda}\Gamma_{\sigma\nu}^\alpha)
-\f{1}{2}\delta_\sigma^\lambda
g^{\mu\nu}\Gamma_{\mu\beta}^\alpha\Gamma_{\nu\alpha}^\beta
=-\varkappa (T_\sigma^\lambda +t_\sigma^\lambda) \, .
\ee

{\bf We will show below how close to the true
gravitational field equations was Einstein when writing the paper of
4 November 1915~[19]. }

Since 1913 A. Einstein mentioned, in one or another way, that the
\textbf{quantity }{\mathversion{bold} \( \!t_\sigma^\lambda \)},
characterizing the
gravitational field must enter the gravitation equation {\bf in the
same
way} as the \textbf{quantity } {\mathversion{bold} \(
\!t_\sigma^\lambda \)}, characterizing material
systems.

For instance, he wrote in 1913 in the paper [8]:
{\it ``...the gravitational field tensor is a source of the field on
equal foots with that of material systems, $\Theta_{\mu\nu}$.
Exceptional position of the gravitational field energy in comparison
with all other kinds of energy would lead to inadmissible
consequences.''}. However, Einstein left aside this important
intuituve argument when he wrote the paper [19]. 

In fact, the mentioned above consideration on a symmetry between the
quantities  \( T_\sigma^\lambda \) and  \(
t_\sigma^\lambda \) is rather a product of Einstein's intuition, but
not a general physical principle. The matter is that the
transformation properties of these quantities are different.

One has to notice that, as a rule, basic physical equations are not
derived.
Rather they are guessed on the basis of experimental data, general
physical principles and intuition. That is why it is sometimes
difficult to logically explain in what way they are obtained by an
author. 

It is easy, with help of (60), to find the trace of the quantity 
\( t_\sigma^\lambda \) 
\be
t=t_\lambda^\lambda=
\f{1}{\kappa}g^{\mu\nu}\Gamma_{\mu\beta}^\alpha
\Gamma_{\nu\alpha}^\beta,
\ee   
and to rewrite Einstein's equation (61) in the form
\be
\pa_\alpha(g^{\nu\lambda}\Gamma_{\sigma\nu}^\alpha)
=-\varkappa \Bigl(T_\sigma^\lambda +t_\sigma^\lambda
-\f{1}{2}\delta_\sigma^\lambda\,t\Bigr).
\ee
It is seen that there is no symmetry between the quantities 
\( T_\sigma^\lambda \) and  \( t_\sigma^\lambda \) in Eq. (63). One
can easily see that this symmetry can be re-established in a simple
way. 

Consider first the conservation laws with help of (63). To this end
we find the trace:
\be
\pa_\alpha(g^{\nu\beta}\Gamma_{\nu\beta}^\alpha)
=-\varkappa (T-t).
\ee
Now we multiply both parts of Eq.(64) by  \(
\ds\f{1}{2}\delta_\sigma^\lambda \)
and subtract the result from (63):
\be
\pa_\alpha\Bigl(g^{\nu\lambda}\Gamma_{\sigma\nu}^\alpha
-\f{1}{2}\delta_\sigma^\lambda\,g^{\nu\beta}\Gamma_{\nu\beta}^\alpha\
Bigr)
=-\varkappa \Bigl(T_\sigma^\lambda +t_\sigma^\lambda
-\f{1}{2}\delta_\sigma^\lambda\,T\Bigr).
\ee
One easily sees that the following equalities hold:
\be
\pa_\lambda \pa_\alpha(g^{\nu\lambda}\Gamma_{\sigma\nu}^\alpha)
=\f{1}{2}\pa_\lambda \pa_\alpha \pa_\sigma g^{\alpha\lambda},
\ee
\be
\pa_\lambda \pa_\alpha\delta_\sigma^\lambda
g^{\nu\beta}\Gamma_{\nu\beta}^\alpha
=\pa_\lambda \pa_\alpha \pa_\sigma g^{\alpha\lambda}.
\ee
Making use of these equalities we find from Eq.(65):
\be
\pa_\lambda (T_\sigma^\lambda +t_\sigma^\lambda)
=\f{1}{2}\delta_\sigma^\lambda \pa_\lambda\,T,
\ee
similarly one can find, using (58), the relation
\be
\pa_\lambda T_\sigma^\lambda
+\f{1}{2}T_{\mu\nu}\pa_\sigma g^{\mu\nu}
=\f{1}{2}\delta_\sigma^\lambda \pa_\lambda T \,.
\ee
It is evident from this that Eq. (63) does not provide the
conservation laws, and also there is no symmetry between 
\( T_\sigma^\lambda \) and  \( t_\sigma^\lambda \).
To re-establish the symmetry in (63) and (68) it is necessary to make
the folowing substitution:
\be
T_\sigma^\lambda \to T_\sigma^\lambda
-\f{1}{2}\delta_\sigma^\lambda T \, ,
\ee  
The trace of the tensor \( T_{\mu\nu} \) is being changed as follows:
\be
T \to -T \, .
\ee  
Note that symmetrization is not related to any assumptions on the
structure of matter. Having completed this operation we obtain the
new gravitational equations
\be
\pa_\alpha (g^{\nu\lambda}\Gamma_{\sigma\nu}^\alpha)
=-\varkappa \bigl\{(T_\sigma^\lambda +t_\sigma^\lambda)
-\f{1}{2}\delta_\sigma^\lambda (T+t)\bigr\}.
\ee
The same operation applied to (68) and (69) leads to re-establishing
of the conservation laws
\be
\pa_\lambda (T_\sigma^\lambda +t_\sigma^\lambda)=0,
\ee
\be
\pa_\lambda T_\sigma^\lambda +\f{1}{2}T_{\mu\nu}\pa_\sigma
g^{\mu\nu}=0 \,
.
\ee
Eqs. (73) and (74) arise only from  new equations (72).

In the supplement [20] to the article [19] Einstein makes a further
step and chooses the gravitational equations in the form
\be R_{\mu\nu}=-\varkappa \,T_{\mu\nu} \, ,
\ee
generally covariant under arbitrary coordinate transformations. He
abandons  the condition 
\( \sqrt{-g}=1. \) In the frame  \( \sqrt{-g}=1 \) these
equations are equivalent to Eq.(52). 
But Eq.(52) does not provide neither  the symmetry between 
\( T_\sigma^\lambda \) and  
\( t_\sigma^\lambda \) 
nor   the conservation laws. So,  it is natural to make the
symmetrisation
operations, 
(70) and (71), in the initial equations (75) as well. In such a way
we obtain a new gravitation equation
\be
R_{\mu\nu}=-\varkappa \left(T_{\mu\nu}-\f{1}{2}g_{\mu\nu}
T\right).
\ee
Namely these equations were  obtained by Einstein several days
later and published then in the paper~[5]. Note that Einstein found
the
conservation law equations (73) still with the gravitation equations
(63). 

This sircumstance, probably, satisfied him, and he did not pay
attention to a  symmetry breaking between 
\( T_\sigma^\lambda \) and  \( t_\sigma^\lambda \) in Eqs.(63).

However his method to satisfy the conservation laws led to the
situation when the choice of the frame 
\( \sqrt{-g}=1 \) was possible only if the trace of the material
tensor were put zero. Instead of re-establishing the symmetry via
(70) and (71) Einstein chose another, more radical, way. He pushed
forward a
new physical idea~[20], that {\it ``in reality only the quantity
\( T_\mu^\mu + t_\mu^\mu \) is positive, while  \( T_\mu^\mu \)
disappears''}. Such an approach re-established the symmetry.
Nonetheless whatever radical, this approach was not fruitful, and
this idea existed but short time. 

Later Einstein returned to his old idea on symmetry and obtained in
Ref.~[5] the gravitational field equations (76). He mentioned
there: \textit{``\textbf{As it is not difficult to see, our
additional
term
leads to that energy tensors of the gravitational field and of matter
enter Eq.(9) in the same way}''.} 

There is some inexactitude in this statement. {\bf There does not
exist,
in general relativity, a gravitational field energy tensor.}

Nonetheless, due to intuitive considerations, the use of such a
quantity led Einstein directly to his goal. 

We see that the way of Einstein led him inevitably to the same
equations, which Hilbert obtained as well. It is quite evident that
Einstein obtained them independently. Moreover, he gained them
through much suffering during several years.

For better understanding of what is written above, of no small
importance is quite a vivid correspondence~[9] between Hilbert anf
Einstein, which took place just in the period of their work on
the gravitational field equations. Namely this correspondence witness
that no \textit{ ``\textbf{radically different}''} point of view,
other
than the
standard one, can exist,  as a matter of principle.

\section*{\Large 3. Einstein--Hilbert Correspondence}

{\bf From Einstein to Hilbert

\hfill Berlin, Sunday, 7 November 1915 }

{``\it  Highly esteemed Colleague,

With return post I am sending  you the correction to a paper in which
I 
changed the gravitational equations, after having myself noticed
about 4 weeks ago that my method of proof was a fallacious one. My
colleague Sommerfeld wrote that you also have found a hair in my
soup that has spoiled it entirely for you. I am curious whether you
will take kindly to this new solution. 

With cordial greetings, yours

\hfill A. Einstein

When may I expect the mechanics and history week to take place in
G\"ottingen? I am looking forward to it very much.''} 

\vspace*{1cm}
\noindent
{\bf From Einstein to Hilbert

\hfill Berlin, Friday, 12 November 1915 }

{``\it  Highly esteemed Colleague,

I just thank you for the time being for your kind letter. The problem
has
meanwhile made new progress. Namely, it is possible to exact
\underline{general} covariance from the postulate $\s{-g}=1$;
Riemann's tensor then delivers the gravitation equations directly. If
my present  modification (which does not change the equations) is
legitimate, then gravitation must play a fundamental role in the
composition of matter. My own curiousity is interfering with my work!
I am sending you two copies of last year's paper. I have only two
other intact copies myself. If comeone else needs the paper, he can
easily purchase one, of course, for 2M (as an Academy offprint). 

Cordial greetings, yours

\hfill Einstein''
}

\vspace*{1cm}
\noindent
{\bf From Hilbert to Einstein

\hfill G\"ottingen, 13 November 1915}

{\it
``Dear Collegue,

Actually, I first wanted to think of a very palpable application for
physicists, namely reliable relations between the physical constants,
before obliging with my axiomatic solution to your great problem. But
since you are so interested, I would like to lay out my theory in
very complete detail on the coming Tuesday, that is, the day after
the day after tomorrow (the 16th of this mo.). I find it ideally
handsome mathematically and absolutely compelling according to
axiomatic method, even to the extent that not quite transparent
calculations do not occur at all and therefore rely on its
factuality. As a result of gen. math. law, the (generalized
Maxwellian) electrody. eqs. as a math. consequence of the gravitation
eqs., such that gravitation and electrodynamics are actually nothing
different at all. Furthermore, my energy concept forms the basis:
$E=\sum (e_st^s+e_{ih}t^{ih})$, which is likewise a general
invariant, and from this then also follow from a very simple axiom
the 4 missing ``space-time equations'' $e_s=0$. I derived most
pleasure in the discovery already duscussed with Sommerfeld that
normal electrical energy results when a specific absolute invariant
is differentiated from the gravitation potentials and then $g$ is set
$=0.1$. My request is thus to come for Tuesday. You can arrive at 3
or $1/2$ past 5. The Math. Soc. meets at 6 o'clock in the auditorium
building. My wife and I would be very pleased if you stayed with us.
It would be better still if you came already on Monday, since we have
the phys. colloquium on Monday, 6 o'clock, at the phys. institute.
With all good wishes and in the hope of soon meeting again, yours,

\hfill Hilbert

As far as I understand your new paper, the solution given by you is
entirely different from mine, especially since my $e_s$'s must also
necessarily contain the electrical potential.
''
}

\vspace*{1cm}
\noindent
{\bf From Einstein to  Hilbert 

\hfill Berlin, Monday, 15 November 1915}

{\it`` Highly esteemed Colleague,

Your analysis interests me tremendously, especially since I often
racked my brains to construct a bridge between gravitation and
electromagnetics. The hints your give in your postcards awaken the
greatest of expectations. Nevertheless, I must refrain from
travelling to G\"ottingen for the moment and rather must wait
patiently untill I can study your system from the printed article;
for I am tired out and plagued with stomach pains besides. If
possible, please send me a correction proof of your study to mitigate
my impatience. 

With best regards and cordial thanks, also to Mrs. Hilbert, yours,

\hfill A. Einstein
''
}

16 November 1915 D. Hilbert presented his result publicly. The
author of the paper [21] writes about that:

{\it  $\ll$
``Grundgleichungen der
Physik'' was the title of Hilbert's lecture to the G\"ottingen
Mathematical Society of November 16. It was also the title under
which his communication in the letter of invitation circulated among
the Academy members between November 15 and the meeting of November
20...
$\gg$
}

He mentions also:

{\it $\ll $The invitation for the meeting of 20 November was issued
on November 15 and was, as always, circulated among the members to
confirm their participation and announce any communications they
intended to present at the meeting. Into this invitation Hilbert
wrote: ``Hilbert legt vor in die Nachrichten: Grundgleichungen der
Physik.''
$\gg$}

\textit{``\textbf{In response to Einstein's request}}'', as the
author of Ref.[21]
notices,
\textit{``\textbf {Hilbert had to report his findings in
correspondence to
Einstein}, unfortunately lost. He probably sent Einstein the
manuscript
of his lecture to the G\"ottingen Mathematical Society, or a summary
of its main points.''}

\newpage
\noindent

{\bf Einstein to  Hilbert

\hfill Berlin, 18 November, 1915}

{\it``Dear Colleague,

The system you furnish agrees --- as far as I can see --- exactly
with what I found in the last few weeks and have presented to the
Academy. The difficulty was not in finding generally covariant
equations for the $g_{\mu\nu}$'s; for this is easily achieved with
the aid of Riemann's tensor. Rather, it was hard to recognize that
these equations are a generalization, that is, simple and natural
generalization of Newton's law. It has just been in the last few
weeks that I succeeded in this (I sent you my communications),
whereas 3 years ago with my friend Grossmann I had already taken into
consideration the only possible generally covariant
equations, which have now been shown to be the correct ones. We had
only heavy-heartedly distanced ourselves from it, because it seemed
to me that the physical discussion yielded an incongruency with
Newton's law. The important thing is that the difficulties have now
been overcome. Today I am presenting to the Academy a paper in which
I derive quantitatively out of general relativity, without any
guiding hypothesis, the perihelion motion of Mercury discovered by Le
Verrier. No gravitation theory had achieved this untill now. 

Best regards, yours

\hfill Einstein
''
}

\vspace*{1cm}

Such is the content of Einstein's reply letter. There does not exist
an argument more forcible than the words in the letter, written by
Einstein, himself: \textit{``\textbf{The system you furnish agrees
--- as far
as I can see --- exactly with what I found in the last few  weeks and
have
presented to the Academy}''}. But namely this exact evidence remained
aside in Refs.~[2,3,4]. Already this only evidence by Einstein 
is
fairly sufficient to exclude completely and forever any attempts to
push forward \textit{``\textbf{a point of view that radically
differs from the
standard point of view}''}. 

The authors of [2,3] made a whole series of other wrong conclusions 
about Hilbert's paper. That is why we had to consider, in Section~1,
their compositions in some detail.

Let us nonetheless assume that Einstein received from Hilbert the
gravitation equations in the form (12), i.e.
\be
[\sqrt{g}\,R]_{\mu\nu}=-\f{\pa \sqrt{g}\,L}{\pa g^{\mu\nu}} \, ,
\ee
It is unbelievable that Einstein would consider that  {\bf these
equations} agreed with {\bf his equations}
\be
R_{\mu\nu}=-\varkappa
\left(T_{\mu\nu}-\f{1}{2}g_{\mu\nu}T\right),
\ee
where the Ricci tensor enters explicitly. To agree that Eqs. (77)
coincide with his equations (78) Einstein would need to calculate the
derivatives
\[
\f{\pa\sqrt{g}\,R}{\pa g^{\mu\nu}},\quad \f{\pa\sqrt{g}\,R}{\pa
g_k^{\mu\nu}},\quad
\f{\pa\sqrt{g}\,R}{\pa g_{k\ell}^{\mu\nu}} \, .
\] 

However he did not calculate them that time. He wrote about that
later, in the letter to H.A.~Lorentz of 19 January 1916~[9]:
{\it ``I avoided the somewhat involved computation of the 
$\partial R/\pa g^{\mu\nu}$'s and  $\pa R/\pa g_\sigma^{\mu\nu} \)'s
by
setting
up the tensor equations directly. But the other way is certainly also
workable and even more elegant mathematically''}.  


{\bf It is also improbable that Hilbert, knowing that Ricci tensor
enters the Einstein equations (he was informed of that in the letter
from Einstein of 7~November 1915), could send him his equations in
the form (77). No doubt that Einstein received from Hilbert the
equations in the form}
\be
\sqrt{g}\left(R_{\mu\nu}-\f{1}{2}g_{\mu\nu}R\right)
=-\f{\pa \sqrt{g}\,L}{\pa g^{\mu\nu}},
\ee
because it was not difficult for Hilbert, to find, from general
considerations and practically without computations, as we have seen
above, the equality
\[
[\sqrt{g}\,R]_{\mu\nu}=\sqrt{g}\left(R_{\mu\nu}-\f{1}{2}g_{\mu\nu}
R\right).
\]

In the letter to Hilbert of 18 November 1915 Einstein wrote:
\textit{``\textbf{The system you furmish agrees --- as far as I can
see ---
exactly to what I found...}''}. It is easy to be persuaded in this if
to compare Eqs.(78) and (79). Einstein's words \textit{``\textbf {as
far as I can
see}''} were possibly caused by that in Hilbert's paper the
energy-momentum tensor density was defined as
\[
\f{\pa \sqrt{g}\,L}{\pa g^{\mu\nu}},
\]
where \( L \)  is a function of  \( g^{\mu\nu},   q_\sigma\) 
and \( q_{\sigma\nu}\). Such a definition was new and unknown to
Einstein. Time needed to understand its essence. But Einstein replied
to Hilbert immediately. Later on, in the paper~[22], Einstein would
take advantage of namely such a definition of the energy-momentum
tensor.
He, in this paper, introduced, like Hilbert, a function 
$\cal{M}$ of variables 
\( g^{\mu\nu}, q_{(\rho)}, q_{(\rho)\alpha} \) and wrote down the
energy-momentum tensor density in the form
\[
{\cal T}_{\mu\nu}=-\frac{\pa {\cal M}} 
{\partial  g^{\mu\nu}}\,.
\]

Therefore it is impossible to understand on what ground the authors
of [3] try to conclude quite an opposite:
{\it ``The new energy expression that Hilbert now took over from
Einstein \ldots''}. 
As we have just seen it is absolutely wrong. Namely Einstein adopted
from Hilbert the definition of the energy-momentum tensor density and
used it in the paper~[22]. 

Furthermore the authors of~[3] conclude:
{\it ``\ldots Einstein's generalization of Hilbert's derivation made
it
possible to regard the latter as merely representing a problematic
special case''}.  

All this is wrong. Hilbert's method is general; it allows to obtain 
the gravitation equation without assumption on a concrete form of the
Lagrange function 
\( L \) of a material system. Therefore there was no (and could not
be) generalization  of the Hilbert inference. This is another story
that
afterwards Hilbert applied his method to the concrete case of Mie's
theory.

As we have already  mentioned in Section~1, the transformation of
(77)
to (79) was not a  great labour for Hilbert with help of Theorem~III,
proven by him. 

So the Proofs, moreover mutilated, cannot witness that Hilbert did
not put the gravitational field equations in the form (79).

\section*{\Large Conclusion}

The analysis, undertaken in Sections~1 and 2, shows that Einstein and
Hilbert inependently discovered the gravitational field equations.
Their pathways were different but they led exactly to the same
result. Nobody ``nostrified'' the other. So no {\it ``belated
decision in the Einstein--Hilbert priority dispute''}, about which
the authors of~[2] wrote, can be taken. Moreover, the very
Einstein--Hilbert dispute never took place.

All is absolutely clear: {\bf both authors made everything to
immortalize
their
names  in the title of  the gravitational field equations}.

But general relativity is  Einstein's theory.


\section*{\Large Acknowledgement}
The authors are indebted to S.S.~Gershtein and N.E.~Tyurin for
valuable discussions of the paper and to C.J.~Bjerknes for helpful
remarks.

\begin{flushright}
\section*{\large Appendix}\addcontentsline{toc}{section}%
{Appendix A}
\end{flushright}

Below we shall give, with pedagogical purposes, the detailed proof of
Hilbert's theorems 
II and  III.\\

\begin{flushleft}
\textbf{Theorem II}
\end{flushleft}
If  \( H \) is an invariant that depends on 
\( g_{\mu\nu}, \pa_\lambda g_{\mu\nu}, \pa_\sigma \pa_\lambda
g_{\mu\nu}, A_\nu\) and \( \pa_\lambda A_\nu \),
then for an infinitesimal contravariant vector 
\( \delta x^s \) the following identity holds:
$$
\delta_L (\sqrt{g}H)=\pa_s(\sqrt{g}H\delta x^s); \eqno(A.1)
$$
here \( \delta_L \) is the Lie variation. 

To prove this theorem consider the integral 
$$
S=\int\limits_\Omega d^4x\sqrt{g}H.\eqno(A.2)
$$
Let us make an infinitesimal coordinate transformation 
$$
x^{\prime\nu}=x^\nu+\delta x^\nu; \eqno(A.3)
$$
here \( \delta x^\nu \) is an arbitrary infinitesimal four-vector. 

At this transformation the integral remains intact and theoref the
variation 
\( \delta_c S \) disappears: 
$$
\delta_c S=\int\limits_{\Omega^\prime} d^4x^\prime
\sqrt{g^\prime}H^\prime
-\int\limits_\Omega d^4x\sqrt{g}H=0.\eqno(A.4)
$$
The first integral may be written as 
$$
\int\limits_{\Omega^\prime} d^4x^\prime \sqrt{g^\prime}H^\prime
=\int\limits_\Omega J\sqrt{g^\prime}H^\prime d^4x.\eqno(A.5)
$$
Here  \( J \) is the Jacobian of the transformation
$$
J=\f{\pa (x^{\prime 0}, x^{\prime 1}, x^{\prime 2}, x^{\prime 3})}
{\pa (x^0, x^1, x^2, x^3)}.\eqno(A.6)
$$
Jacobian of the transformations \( (A.3) \) is 
$$
J=1+\pa_\lambda \delta x^\lambda .\eqno(A.7)
$$
Expanding  \( \sqrt{g^\prime}H^\prime \) into the Taylor series one
finds 
$$
\sqrt{g^\prime (x^\prime)}H^\prime (x^\prime)
=\sqrt{g^\prime (x)}H^\prime (x)
+\delta x^\lambda \pa_\lambda (\sqrt{g}H).\eqno(A.8)
$$
Due to  \( (A.5), (A.7) \) and  \( (A.8) \)
equality   \( (A.4) \) assumes the form:
$$
\delta_c S=\int\limits_\Omega d^4x[\delta_L(\sqrt{g}H)
+\pa_\lambda (\sqrt{g}H\delta x^\lambda)]=0.\eqno(A.9)
$$
The Lie variation is 
$$
\delta_L(\sqrt{g}H)
=\sqrt{g^\prime (x)}H^\prime (x)
-\sqrt{g(x)}H(x).\eqno(A.10)
$$
The Lie variation commutes with partial derivatives: 
$$
\delta_L \pa_\lambda 
=\pa_\lambda \delta_L .\eqno(A.11) 
$$
The Lie variation of \( \sqrt{g}H \) is 
$$
\delta_L(\sqrt{g}H)
=P_g(\sqrt{g}H)
+P_q(\sqrt{g}H),\eqno(A.12)
$$
where 
$$
\!P_g(\sqrt{g}H)
=\f{\pa \sqrt{g}H}{\pa g_{\mu\nu}}\delta_L g_{\mu\nu}
+\f{\pa \sqrt{g}H}{\pa (\pa_\lambda g_{\mu\nu})}\pa_\lambda\delta_L
g_{\mu\nu}
+\f{\pa \sqrt{g}H}
{\pa (\pa_\sigma \pa_\lambda g_{\mu\nu})}\pa_\sigma \pa_\lambda 
\delta_L g_{\mu\nu},\eqno(A.13)
$$
$$
P_q(\sqrt{g}H)
=\f{\pa \sqrt{g}H}{\pa A_\lambda}\delta_L A_\lambda
+\f{\pa \sqrt{g}H}
{\pa (\pa_\sigma A_\lambda)}\pa_\sigma \delta_L A_\lambda.\eqno(A.14)
$$
Due to arbitrariness of the volume  \( \Omega \) one gets on the
basis of  \( (A.9) \)
the sesired Hilbert identity: 
$$
\delta_L (\sqrt{g}H)
+\pa_\lambda (\sqrt{g}H\delta x^\lambda)\equiv 0,\eqno(A.15)
$$
where 
$$
\delta_L (\sqrt{g}H)
=P_g (\sqrt{g}H)
+P_q (\sqrt{g}H).\eqno(A.16)
$$
\begin{flushleft}
\textbf{Theorem III}
\end{flushleft}
If an invariant depends on 
\( g_{\mu\nu}, \pa_\lambda g_{\mu\nu}, \pa_\sigma 
\pa_\lambda g_{\mu\nu}\) then the variational derivative 
$$
\f{\delta \sqrt{g}H}{\delta g_{\mu\nu}}
=\sqrt{g}G^{\mu\nu} 
=\f{\pa \sqrt{g}H}{\pa g_{\mu\nu}}
-\pa_\lambda \f{\pa \sqrt{g}H}{\pa (\pa_\lambda g_{\mu\nu})}
+\pa_\sigma \pa_\lambda \f{\pa \sqrt{g}H}{\pa (\pa_\sigma \pa_\lambda
g_{\mu\nu})}\eqno(A.17)
$$
satisfies the identity 
$$
\nabla_\lambda G^{\lambda\nu}\equiv 0,\eqno(A.18)
$$
or, in another form, 
$$
\pa_\lambda (\sqrt{g}G_\rho^\lambda) 
+\f{1}{2}\s g G_{\lambda\sigma}\pa_\rho g^{\lambda\sigma}\equiv
0,\eqno(A.19)
$$
where  \( \nabla_\lambda \) stands for the covariant derivative in
the
Riemann space.

To prove this theorem consider the integral 
$$
\int\limits_\Omega\sqrt{g}H d^4x
$$
over a finite region of the four-dimensional world. 
The translation vector \( \delta x^\sigma \) 
in  \( (A.3) \) has to disappear together with its derivatives on the 
3-dimensional border of the region  \(\Omega \).
This implies dissapearing of the field variations and their
derivatives on the border of this region. Taking use of the Hilbert
identity 
\( (A.15) \) one finds: 
$$
\int\limits_\Omega \delta_L (\sqrt{g}H)d^4x=0. \eqno(A.20)
$$
In our case 
$$
\delta_L (\sqrt{g}H)=P_g(\sqrt{g}H). \eqno(A.21)
$$
Expression  \( (A.13) \) can be written in the form 
$$
P_g(\sqrt{g}H)
=\f{\delta \sqrt{g}H}{\delta g_{\mu\nu}}
\delta_L g_{\mu\nu}+\pa_\lambda S^\lambda, \eqno(A.22)
$$
where vector  \( S^\lambda \) is 
$$
S^\lambda 
=\left[\f{\pa \sqrt{g}H}{\pa (\pa_\lambda g_{\mu\nu})}
-\pa_\sigma \left(\f{\pa \sqrt{g}H}{\pa (\pa_\sigma \pa_\lambda
g_{\mu\nu})} \right) \right]\delta_L g_{\mu\nu}
+\f{\pa \sqrt{g}H}{\pa (\pa_\sigma \pa_\lambda g_{\mu\nu})}
\pa_\sigma \delta_L g_{\mu\nu}.\eqno(A.23)
$$
Substituting  \( (A.22)\) into   \( (A.20) \) one finds: 
$$
\int\limits_\Omega \f{\delta \sqrt{g}H}{\delta g_{\mu\nu}}
\delta_L g_{\mu\nu} d^4x=0. \eqno(A.24)
$$

Now we shall find the variation  \( \delta_L g_{\mu\nu} \)
at the transformations  \( (A.3) \). Metric tensor   \( g_{\mu\nu} \)
is transformed as 
$$
g_{\mu\nu}^\prime (x^\prime)
=\f{\pa x^\lambda}{\pa x^{\prime \mu}}\cdot
\f{\pa x^\sigma}{\pa x^{\prime \nu}}g_{\lambda\sigma}(x).
$$
One finds thereof for the transformation  \( (A.3) \)
$$
\delta_L g_{\mu\nu}(x)
=-\delta x^\sigma \pa_\sigma g_{\mu\nu}
-g_{\mu\sigma}\pa_\nu \delta x^\sigma
-g_{\nu\sigma}\pa_\mu \delta x^\sigma. \eqno(A.25)
$$
With account of the equality 
$$
\nabla_\sigma g_{\mu\nu}
=\pa_\sigma g_{\mu\nu}
-g_{\lambda\mu}\varGamma_{\sigma\nu}^\lambda
-g_{\lambda\nu}\varGamma_{\sigma\mu}^\lambda =0,\eqno(A.26)
$$
one can write the Lie derivative in the covariant form: 
$$
\delta_L g_{\mu\nu}
=-g_{\mu\sigma}\nabla_\nu \delta x^\sigma
-g_{\nu\sigma}\nabla_\mu \delta x^\sigma.\eqno(A.27)
$$
Substituting this expression into the integral \( (A.24) \) one gets 
$$
\int\limits_\Omega d^4x\f{\delta\sqrt{g}H}{\delta g_{\mu\nu}}
g_{\mu\sigma}\nabla_\nu \delta x^\sigma =0.\eqno(A.28)
$$
Eq.  \( (A.28) \) can be written in the form 
$$
\int\limits_\Omega \left[\nabla_\nu \left(
\f{\delta\sqrt{g}H}{\delta g_{\mu\nu}}g_{\mu\sigma}\delta x^\sigma
\right)
-\delta x^\sigma \nabla_\nu 
\left(\f{\delta\sqrt{g}H}{\delta g_{\mu\nu}}g_{\mu\sigma}\right)
\right]d^4x=0.\eqno(A.29)
$$
Note, that 
$$
\nabla_\nu \left(\f{\delta\sqrt{g}H}{\delta g_{\mu\nu}}
g_{\mu\sigma}\delta x^\sigma \right)
=\pa_\nu \left(\f{\delta\sqrt{g}H}{\delta g_{\mu\nu}}
g_{\mu\sigma}\delta x^\sigma\right).\eqno(A.30)
$$
Due to  \( (A.30) \) the integral of the first term in the l.h.s. of  
\( (A.29) \) disappears and Eq.  \( (A.29) \) assumes the form:
$$
\int\limits_\Omega \delta x^\sigma \nabla_\nu G_\sigma^\nu
d^4x=0.\eqno(A.31)
$$
Here we have introduced in accordance with definition  \( (A.17) \)
the mixed tensor: 
\[
\sqrt{g}G_\sigma^\nu
=\f{\delta \sqrt{g}H}
{\delta g_{\mu\nu}}g_{\mu\sigma}.
\]
We find, due to arbitrariness of the vector  \( \delta x^\sigma \) 
the desired Hilbert identity 
$$
\nabla_\nu G_\sigma^\nu\equiv 0.\eqno(A.32)
$$
or, in more detail, 
$$
\nabla_\nu G_\sigma^\nu
=\pa_\nu G_\sigma^\nu
-\varGamma_{\sigma\nu}^\lambda G_\lambda^\nu
+\varGamma_{\nu\lambda}^\nu G_\sigma^\lambda\equiv 0.\eqno(A.33)
$$
With account of the expression 
$$
\varGamma_{\sigma\nu}^\lambda 
=\f{1}{2}g^{\lambda\rho}
(\pa_\sigma g_{\nu\rho}+\pa_\nu g_{\sigma\rho}-\pa_\rho
g_{\sigma\nu}),\quad
\pa_\lambda \sqrt{g}=\sqrt{g}\varGamma_{\nu\lambda}^\nu,\eqno(A.34)
$$
one finds 
$$
\nabla_\nu (\sqrt{g}G_\rho^\nu)
=\pa_\nu (\sqrt{g}G_\rho^\nu)
+\f{1}{2}\sqrt{g}G_{\lambda\sigma}\pa_\rho
g^{\lambda\sigma}=0.\eqno(A.35)
$$
This identity was obtained by Hilbert in 1915. 

Applying this identity to the invariant \( H=R \), where  \( R \) 
is the scalar curvature, D.~Hilbert obtained the Bianchi idenrity 
$$
\nabla_\nu (R^{\mu\nu}-\f{1}{2}g^{\mu\nu}R)\equiv 0.\eqno(A.36)
$$
The detailed account of that is in the main text of this article.  

Now we apply Theorem  II to the invariant  \( L \), which depends on
\( A_\nu, \pa_\lambda A_\nu,\break  g_{\mu\nu},\pa_\lambda g_{\mu\nu}
\).
One has on the basis of \( (A.22) \) 
$$
P_g(\sqrt{g}L)
=\f{\delta \sqrt{g}L}{\delta g_{\mu\nu}}
\delta_L g_{\mu\nu}+\pa_\lambda S_1^\lambda,\eqno(A.37)
$$
where 
$$
S_1^\lambda
=\f{\pa \sqrt{g}L}{\pa(\pa_\lambda g_{\mu\nu})}\delta_L
g_{\mu\nu}.\eqno(A.38)
$$
Likewise 
$$
P_q(\sqrt{g}L)
=\f{\delta \sqrt{g}L}{\delta A_\lambda}\delta_L A_\lambda 
+\pa_\lambda S_2^\lambda,\eqno(A.39)
$$
where 
$$
S_2^\lambda
=\f{\pa \sqrt{g}L}{\pa(\pa_\lambda A_\sigma)}
\delta_L A_\sigma.\eqno(A.40)
$$
One finds from  \( (A.15) \), \( (A.37) \) and  \( (A.39) \) 
$$
\int\limits_\Omega
\left[\f{\delta \sqrt{g}L}{\delta g_{\mu\nu}}\delta_L g_{\mu\nu}
+\f{\delta \sqrt{g}L}{\delta A_\lambda}\delta_L A_\lambda \right]
d^4x=0.\eqno(A.41)
$$ 

Now let us find the Lie variation of the field variable  \( A_\lambda
\).
According to the transformation law for the vector \( A_\lambda \) we
get 
$$
A_\lambda^\prime (x^\prime) 
=\f{\pa x^\nu}{\pa x^{\prime \lambda}}A_\nu (x).\eqno(A.42)
$$
Thereof we find for transformation \( (A.3) \) 
$$
A_\lambda^\prime (x+\delta x) 
=A_\lambda (x)-A_\nu (x)\pa_\lambda \delta x^\nu.\eqno(A.43)
$$
Expanding the l.h.s. into the Taylor series we obtain 
$$
\delta_L A_\lambda 
=A_\lambda^\prime (x)
-A_\lambda (x)
=-\delta x^\nu \pa_\nu A_\lambda
-A_\nu (x)\pa_\lambda \delta  x^\nu,\eqno(A.44)
$$
or, in the covariant form, 
$$
\delta_L A_\lambda 
=-\delta x^\sigma \nabla_\sigma A_\lambda
-A_\sigma \nabla_\lambda \delta  x^\sigma.\eqno(A.45)
$$
Substituting  \( (A.27) \) and   \( (A.45) \) into  \( (A.41) \) we
find 
$$
\!\!\!\!\!\int\limits_\Omega d^4x\left[2\nabla_\nu \left(
\f{\delta\sqrt{g}L}{\delta g_{\mu\nu}}g_{\mu\sigma}\right)
-\f{\delta\sqrt{g}L}{\delta A_\lambda}\nabla_\sigma A_\lambda
+\nabla_\lambda\left(\f{\delta\sqrt{g}L}{\delta
A_\lambda}A_\sigma\right)
\right]\delta x^\sigma=0.\eqno(A.46)
$$
Due to arbitrariness of the transformation vector  \( \delta x^\sigma
\)
we obtain an identity: 
$$
\!2\nabla_\nu \left(
\f{\delta\sqrt{g}L}{\delta g_{\mu\nu}}g_{\mu\sigma}\right)
=(\nabla_\sigma A_\lambda -\nabla_\lambda A_\sigma)
\f{\delta\sqrt{g}L}{\delta A_\lambda}
-A_\sigma \nabla_\lambda
\left(\f{\delta\sqrt{g}L}{\delta A_\lambda}\right).\eqno(A.47)
$$
According to Hilbert the energy-momentum tensor density is defined by
the expression
$$
T^{\mu\nu} 
=-2\f{\delta\sqrt{g}L}{\delta g_{\mu\nu}}.\eqno(A.48)
$$
Identity \( (A.47) \) assumes the form: 
$$
\nabla_\nu T_\sigma^\nu
=A_\sigma\nabla_\lambda\left(
\f{\delta\sqrt{g}L}{\delta A_\lambda}\right)
+(\nabla_\lambda A_\sigma -\nabla_\sigma A_\lambda)
\f{\delta\sqrt{g}L}{\delta A_\lambda},\eqno(A.49)
$$
or 
$$
\nabla_\nu T_\sigma^\nu
=A_\sigma\pa_\lambda\left(
\f{\delta\sqrt{g}L}{\delta A_\lambda}\right)
+(\pa_\lambda A_\sigma -\pa_\sigma A_\lambda)
\f{\delta\sqrt{g}L}{\delta A_\lambda}.\eqno(A.50)
$$
When the gravitation equations hold, Theorem  III leads to the
equality 
$$
\nabla_\nu T_\sigma^\nu =0,\eqno(A.51)
$$
and, hence, identity \( (A.50) \) transforms into the equation,  
which Hilbert designated  as Eq.(27) in [6]: 
$$
A_\sigma\pa_\lambda\left(
\f{\delta\sqrt{g}L}{\delta A_\lambda}\right)
+(\pa_\lambda A_\sigma -\pa_\sigma A_\lambda)
\f{\delta\sqrt{g}L}{\delta A_\lambda}=0.\eqno(A.52)
$$
But this equation holds always due to Hilbert's Axiom~I, because 
$$
\f{\delta\sqrt{g}L}{\delta A_\lambda}=0.\eqno(A.53)
$$

\ed